\documentclass[preprint,aps,12pt,preprintnumbers,eqsecnum,nofootinbib,superscriptaddress]{revtex4}
\usepackage{graphicx}
\usepackage{amssymb,amsmath}

\def\be{\begin{equation}}
\def\ee{\end{equation}}
\def\beq{\begin{eqnarray}}
\def\eeq{\end{eqnarray}}

\begin{document}
%
%
\title{\vspace*{0.5in} Decaying Dark Matter from Dark Instantons
\vskip 0.1in}
\author{Christopher D. Carone}\email[]{cdcaro@wm.edu}
\author{Joshua Erlich}\email[]{jxerli@wm.edu}
\author{Reinard Primulando}\email[]{rprimulando@email.wm.edu}
\affiliation{Particle Theory Group, Department of Physics,
College of William and Mary, Williamsburg, VA 23187-8795}
\date{August 2010}
\begin{abstract}
We construct an explicit, TeV-scale model of decaying dark matter in which the approximate stability of the
dark matter candidate is a consequence of a global symmetry that is broken only by instanton-induced
operators generated by a non-Abelian dark gauge group.  The dominant dark matter decay channels are to
standard model leptons. Annihilation of the dark matter to standard model states occurs primarily through
the Higgs portal.  We show that the mass and lifetime of the dark matter candidate in this model
can be chosen to be consistent with the values favored by fits to data from the PAMELA and Fermi LAT
experiments.
\end{abstract}
\maketitle

\section{Introduction} \label{sec:intro}
Evidence has been accumulating for an electron and positron excess in cosmic rays compared with expectations
from known galactic sources.  Fermi LAT~\cite{Abdo:2009zk} and H.E.S.S.~\cite{Aharonian:2009ah} have measured
an excess in the flux of electrons and positrons up to a TeV or more.  The PAMELA satellite is sensitive to
electrons and positrons up to a few hundred GeV in energy, and is able to distinguish positrons from electrons
and charged hadrons. PAMELA detects an upturn in the fraction of positron events beginning around
7 GeV~\cite{Adriani:2008zr}.  This is in contrast to the expected decline in the positron fraction from
secondary production mechanisms. Curiously, no corresponding excess of protons or antiprotons has been
detected~\cite{Adriani:2008zq}.

Although conventional astrophysical sources may ultimately prove the explanation of the
anomalous cosmic ray data~\cite{astro}, an intriguing possibility is that dark matter annihilation or decay
provides the source of the excess leptons. If dark matter annihilation is responsible for the excess
leptons, then the annihilation cross section typically requires a large boost factor $\sim 100-1000$ to
produce the observed signal~\cite{Cholis:2008hb}. Possible sources of the boost factor include Sommerfeld
enchancement from additional attractive interactions in the dark sector~\cite{ArkaniHamed:2008qn},
WIMP capture~\cite{MarchRussell:2008tu,Pospelov:2008jd} or Breit-Wigner resonant
enhancement~\cite{feldman,Ibe:2008ye,Guo:2009aj}.

Alternatively, decaying dark matter can provide an explanation of the cosmic ray data if the dark
matter decay channels favor leptonic over hadronic final states~\cite{ddecay}.  A typical scenario of this 
type that is consistent with PAMELA and Fermi LAT data includes dark matter with a mass of a few TeV 
that decays to leptons, with an anomalously long lifetime of $\sim 10^{26}$ 
seconds~\cite{Ibarra:2008jk,Ibarra:2009dr}.  From a model-building perspective, an intriguing issue 
is the origin of this long lifetime, and whether it can be explained with a minimum of theoretical 
contrivance.  With this goal in mind, we present a new model of TeV-scale dark matter, one in which 
an anomalous global symmetry prevents dark matter decays except through instantons of a non-Abelian 
gauge field in the dark sector.  Instanton-induced decays naturally produce the long required lifetime.  
Small mixings between standard model leptons and dark fermions gives rise to the leptonic final states 
observed in the cosmic ray data. Dark matter annihilation through the Higgs portal allows for the 
appropriate dark matter relic abundance, with dark matter masses consistent with the range preferred 
by PAMELA and Fermi-LAT data.

Superheavy dark matter decays through instantons have been considered before as a possible explanation
for ultra-high energy cosmic ray signals, but those scenarios assumed superheavy dark matter with a mass
of $10^{13}$~GeV or higher~\cite{Kuzmin:1997cm} which cannot simultaneously explain the lower energy
electron and positron flux being considered here.  Models of anomaly-induced dark matter decays
without a dark gauge sector can also be constructed.  For example, a supersymmetric extension of the
radiative seesaw model of neutrino masses can explain the PAMELA data through dark matter decays via
an anomalous discrete symmetry~\cite{kubo}. The TeV-scale model we present, which is based on the smallest, 
continuous non-Abelian dark gauge group and smallest set of exotic particles necessary to implement our 
idea, suggests a prototypical set of new particles and interactions that could perhaps be probed at the LHC.

In Section~\ref{sec:model} we present the model and describe the leptonic decay mode via instantons.  In
Section~\ref{sec:relicdensity} we consider dark matter annihilation channels and demonstrate that annihilation
through the Higgs portal can lead to the measured dark matter relic density. In Section~\ref{sec:direct} we
consider dark matter interactions with nuclei and find that our model is safely below current direct detection bounds.
We conclude in Section~\ref{sec:conc}.

\section{The Model} \label{sec:model}
\begin{figure}
    \centering
        \includegraphics[width=8cm,angle=0]{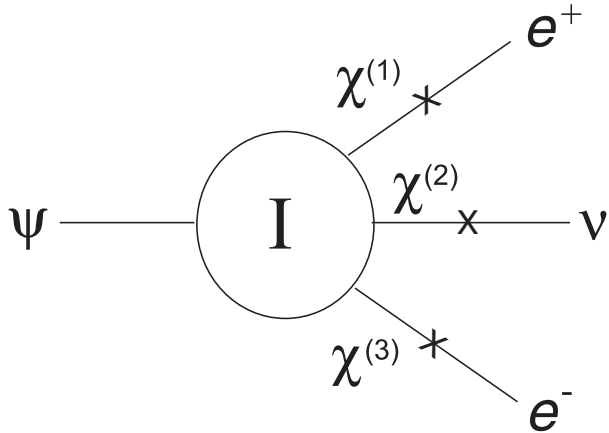}
    \caption{Dark matter decay vertex. The circle represents the instanton-induced
     interaction, while X's represent mass mixing between the $\chi$ fields and
     standard model leptons. Note that $e$ and $\nu$ represent leptons of any generation.}
    \label{fig:one}
\end{figure}
The gauge group of the dark sector is SU(2)$_D\times$U(1)$_D$.  The matter content
consists of four sets of left-handed SU(2)$_D$ doublets and right-handed singlets:
\begin{eqnarray}
&& \psi_L \equiv \left(\begin{array}{c} \psi_u \\ \psi_d \end{array}\right)_L \,\,
\psi_{uR},\,\, \psi_{dR} \,\, ; \,\,\,\,\,
\chi_L^{(i)}\equiv \left(\begin{array}{c} \chi^{(i)}_u \\ \chi^{(i)}_d \end{array}\right)_L \,\,
\chi_{uR}^{(i)}, \,\, \chi_{dR}^{(i)}  \,\,\,\,(i=1 \ldots 3)
\end{eqnarray}
We include an SU(2)$_D$ doublet and singlet Higgs field, $H_D$ and $\eta$, respectively,
that are responsible for completely breaking the dark gauge group.  In addition,
the Higgs field $H_D$ is responsible for giving Dirac masses to the $\psi$ and $\chi$
fields.  The model is constructed so that $\psi$ number corresponds to
an anomalous global symmetry that is violated by the $\psi\chi^{(1)}\chi^{(2)}\chi^{(3)}$
vertex generated via SU(2)$_D$ instantons, as indicated in Fig.~\ref{fig:one}.  The
$\chi$ fields are assigned hypercharges so that they mix with standard model
leptons, leading to the decay $\psi \rightarrow \ell^+ \ell^- \nu$.  The required
lifetime ($\sim \!10^{26}$~s) and the appropriate dark matter relic density
($\Omega_D h^2 \sim 0.1$) constrain the free parameters of the model.

The charge assignments for these fields are summarized in Table~\ref{table:one}.
\begin{table}
\begin{tabular}{cccccc} \hline \hline
$\psi_L$ & $({\bf 2},-1/2)_0$ & \qquad &\qquad &  $\psi_{uR} , \, \psi_{dR}$ & $({\bf 1},-1/2)_0$ \\
$\chi_L^{(1)}$ & $({\bf 2},+1/6)_+$ & \qquad &\qquad &  $\chi^{(1)}_{uR} , \,
\chi^{(1)}_{dR}$ & $({\bf 1},+1/6)_+$ \\
$\chi_L^{(2)}$ & $({\bf 2},+1/6)_0$ & \qquad & \qquad & $\chi^{(2)}_{uR} , \,
\chi^{(2)}_{dR}$ & $({\bf 1},+1/6)_0$ \\
$\chi_L^{(3)}$ & $({\bf 2},+1/6)_-$ & \qquad &\qquad &  $\chi^{(3)}_{uR} , \,
\chi^{(3)}_{dR}$ & $({\bf 1},+1/6)_-$ \\ \hline
$H_D$ & $({\bf 2},0)_0$ & \qquad & \qquad &  $\eta$ & $({\bf 1},1/6)_0$ \\ \hline\hline
\end{tabular}
\caption{Particles charged under the dark gauge groups.  The SU(2)$_D\times$U(1)$_D$ charge
assignments are indicated in parentheses; the subscripts $+$, $-$ and $0$ represent
the standard model hypercharges $+1$, $-1$ and $0$, respectively. Note that the $\psi$ and
$\chi$ states are fermions, while the $H_D$ and $\eta$ are complex scalars.} \label{table:one}
\end{table}
Let us first discuss the consistency of the charge assignments. Cancellation of the
SU(2)$_D^2$ U(1) anomalies requires that the sum of the U(1) charges over all the dark doublet
fermion fields must vanish.  As one can see from Table~\ref{table:one}, this is clearly the
case for the U(1)$_D$ and U(1)$_Y$ charges of the left-handed doublet $\psi$ and $\chi$ fields.
Since SU(2) is an anomaly free group and has traceless generators, all other SU(2)$_D$ anomalies
vanish trivially.  Now consider the U(1)$_D^p$U(1)$_Y^q$ anomalies (where $p$ and $q$ are non-negative
integers satisfying $p+q=3$).  For each field in Table~\ref{table:one} with a given U(1)$_D \times$U(1)$_Y$
charge assignment, one notes that there is another with the same charge assignment but opposite
chirality.  As far as the Abelian groups are concerned, the theory is vector-like and the corresponding
anomalies vanish. Finally, we note that the theory has precisely four SU(2)$_D$ doublets and is
free of a Witten anomaly.

The gauge symmetries of the model lead to a global U(1)$_\psi$ symmetry that prevents the decay of the
lightest $\psi$ mass eigenstate at any order in perturbation theory.  To confirm this statement, we need to
show that all renormalizable interactions that violate this symmetry are forbidden by the dark-sector gauge
symmetry.  The possible problematic interactions that could violate this global symmetry fall into the
following categories:

1. Terms involving $\overline{\psi^c}\psi$.  Here the superscript indicates charge conjugation,
$\psi^c \equiv i \gamma^0\gamma^2\overline{\psi}^T$. This combination has U(1)$_\psi$ charge $+2$.  However, it also
has U(1)$_D$ charge $-1$.  Since we have no Higgs field with the U(1)$_D$ charge $\pm 1$, there are no
renormalizable interactions that violate $\psi$ number by two units.

2. Terms involving a $\chi$ fermion and $\psi$ or $\psi^c$.  Such terms violate $\psi$ number by $\pm 1$
unit.  However, the possible bilinears involving $\psi$ and any $\chi$ have U(1)$_D$ charges $\pm 1/3$
or $\pm 2/3$.  Again, we have no Higgs field with the necessary U(1)$_D$ charge to form a
renormalizable gauge invariant term of this type.

3. Terms involving a standard model fermion and $\psi$ or $\psi^c$.  Such an interaction would violate
$\psi$ number by $\pm 1$, but would have U(1)$_D$ charge $\pm 1/2$.  Again, we have no Higgs fields with
charge $\pm 1/2$ that would allow the construction of a renormalizable invariant.

Since the renormalizable interactions of the theory have an unbroken U(1)$_\psi$ symmetry, no perturbative
process involving these interactions will violate the global symmetry.  However, since the
SU(2)$_D^2$ U(1)$_\psi$ anomaly is non-zero, non-perturbative interactions due to instantons will generate
operators that violate the U(1)$_\psi$ symmetry.

Instantons are gauge field configurations which stationarize the Euclidean action but have a nontrivial
winding number around the three-sphere at infinity.
Following 't Hooft~\cite{thooft1,thooft2}, if there are $N_f$ Dirac pairs of chiral fermions which transform
in the fundamental representation of a gauge group, then  due to the chiral anomaly
a one-instanton configuration violates the
axial U(1)$_A$ charge by 2$N_f$ units.  The non-Abelian, SU$(N_f)\times$SU$(N_f)$ chiral
symmetry is non-anomalous, so the instanton process must involve the 2$N_f$ chiral fermions in a
symmetric fashion.  Fig.~\ref{fig:one} shows the effective
$\psi\chi^{(1)}\chi^{(2)}\chi^{(3)}$ interaction induced by the instanton configuration in our model.\footnote{
In this model, Planck-suppressed operators of this form, if they are present, are negligible compared to the
instanton-induced effects.}  Given the hypercharge assignments
of the $\chi$ fields, these states have electric charges $+1$, $0$ and $-1$, the same as standard model
leptons, of any generation.  After the dark and standard model gauge symmetries are spontaneously broken, there
is no symmetry which prevents the $\chi$ states and the standard model leptons from mixing.  By including a single
vector-like lepton pair, we now show that mixing leading to the decay $\psi \rightarrow \ell^+ \ell^- \nu$
can arise via purely renormalizable interactions.

\begin{figure}
    \centering
        \includegraphics[width=8cm,angle=0]{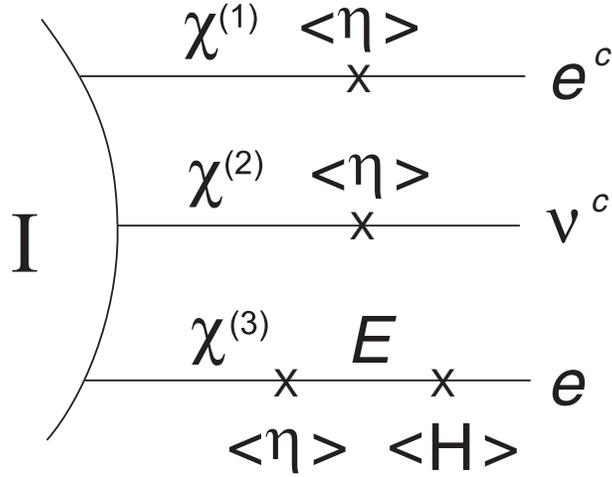}
    \caption{Diagrammatic interpretation of mixing from $\chi$ states to standard model fermions,
corresponding to the right-hand-side of Fig.~\ref{fig:one}. Here $E$ represents the vector-like
lepton described in the text, and $H$ is the standard model Higgs.}
    \label{fig:two}
\end{figure}

We introduce a vector-like lepton pair, $E_L$, $E_R$, with mass $M_E$ and the same quantum numbers
as a right-handed electron; in the notation of Table~\ref{table:one}:
\begin{equation}
E_L \sim E_R \sim ({\bf 1}, 0)_- \,\,\,.
\end{equation}
In addition, we assume in this model that standard model neutrinos have purely Dirac masses. If the Higgs
vacuum expectation values (vevs)
are smaller than the masses of the heavy states, then the mixing to standard model leptons shown in
Fig.~\ref{fig:one} can be estimated via the diagram in Fig.~\ref{fig:two}. Otherwise, one has to diagonalize
the appropriate fermion mass matrices. We discuss the exact diagonalization in an appendix for the reader
who is interested in the details.  Here, the diagrammatic approach is sufficient to establish that the mixing
is present, and is no larger than order $\langle \eta \rangle/ M_\chi$, $\langle \eta \rangle/ M_\chi$, and
$\langle \eta \rangle \langle H \rangle/ (M_\chi M_E)$, where $H$ is the standard model Higgs,
for the $\chi^{(1)}_L-e_R^c$, $\chi^{(2)}_L-\nu_R^c$ and $\chi^{(3)}_L-e_L$ mixing angles, respectively. We
take each mixing angle to be $0.01$ in the estimates that follow, and demonstrate in the appendix how this
choice can be easily obtained.  Further, we assume that decays to the heavy eigenstates are not kinematically
allowed, as is also illustrated in the appendix. Due to the mixing, the $\chi^{(i)}$ particles decay quickly
to standard model particles via couplings to the Higgs bosons and standard model electroweak gauge bosons. The
heavier $\psi$ mass eigenstate decays to lighter states via SU(2)$_D$ gauge-boson-exchange interactions.

The instanton-induced vertex in Fig.~\ref{fig:one} follows from an interaction of the form
\begin{eqnarray}
{\cal L}_I &=& \frac{C}{6 \, g_D^8} \exp\left(-\frac{8 \pi^2}{g_D^2}\right) \left(\frac{m_\psi}{v_D}\right)^{35/6} 
\frac{1}{v_D^2} \,(2 \,\delta_{\alpha\beta} \delta_{\gamma\sigma}
- \delta_{\alpha\sigma} \delta_{\beta\gamma})
\nonumber \\
&& \,\,\,\,\, \cdot \left[ (\overline{\chi^{(2)\,c}_{L\,\beta}} \psi_{L}^\alpha)(\overline{\chi_{L\,\sigma}^{(1)\,c}} 
\chi^{(3)\,\gamma}_{L}) -(\overline{\chi^{(1)\,c}_{L\,\beta}} \psi_{L}^\alpha)(\overline{\chi_{L
\,\sigma}^{(2)\,c}} \chi^{(3)\,\gamma}_{L})
\right] + \mbox{ h.c.} \, ,
\label{eq:od6}
\end{eqnarray}
where $\alpha$, $\beta$, $\gamma$ and $\sigma$ are SU(2)$_D$ indices~\cite{thooft2,noble}.
The dimensionless coefficient $C$ can be computed using the results in Ref.~\cite{thooft2} and one 
finds $C \approx 7 \times 10^8$.  The operators in Eq.~(\ref{eq:od6}) lead, via mixing, to operators of 
the form $\bar \nu_R \psi_L \bar e_R e_L$ and $\bar e_R \psi_L \bar \nu_R e_L$.  Assuming that the product 
of mixing angles is $\approx 10^{-6}$, as discussed earlier, one may estimate the decay width:
\begin{equation}
\Gamma (\psi \rightarrow \ell^+ \ell^- \nu) \approx
\frac{1}{g_D^{16}} \exp(-16 \pi^2/g_D^2) \left(\frac{m_\psi}{v_D}\right)^{47/3} m_\psi \,\,\,.
\end{equation}
For example, for $m_\psi = 3.5$~TeV and $v_D = 4$~TeV, one obtains a dark matter lifetime of $10^{26}$~s
for
\begin{equation}
g_D \approx 1.15 \,,
\end{equation}
where $g_D$ is defined in dimensional regularization and renormalized at the scale 
$m_\psi$~\cite{thooft2}. For similar parameter choices, one can slightly adjust $g_D$ to maintain 
the desired lifetime. As mentioned earlier, dimension-six Planck-suppressed operators are much smaller 
than the operators in Eq.~(\ref{eq:od6}).  Sphaleron-induced interactions are suppressed 
by $\sim \exp[-4 \pi v_D/(g_D T)] \sim \exp(-44 \mbox{ TeV}/\,T)$, and become negligible well before 
the temperature at which dark matter freeze out occurs. 

Finally, let us consider whether the choice $v_D=4$~TeV conflicts with
other meaningful constraints on the heavy particle content of the model.  In short, 
a spectrum of $\sim 4$~TeV $\chi$ and $E$ fermions with order $0.01$ mixing angles with 
standard model leptons presents no phenomenological problems.  These states are above
all direct detection bounds; they are vector-like under the standard model gauge group
so that the $S$ parameter is small; they mix weakly enough with standard model leptons
so that other precision observables are negligibly affected.  On this last point, we note
that the correction to the muon and $Z$-boson decay widths due to the fermion mixing is a
factor of $10^{-8}$ smaller than the widths predicted in the standard model, which is 
within the current experimental uncertainties.  The dark sector gauge bosons are also 
phenomenologically safe.  They do not have couplings that distinguish standard model lepton 
flavor (since they do not couple directly to standard model leptons) so that tree-level 
lepton-flavor violating processes are absent.  The effective four-standard-model-fermion 
operators that are induced by dark gauge boson exchanges are suppressed by 
$\sim (0.01)^4/ v_D^2 \sim 1/(40,000\mbox{ TeV})^2$, which is consistent with the existing 
contact interaction bounds~\cite{pdg}.

We now turn to the question of whether the model provides for the appropriate
dark matter relic density.

\section{Relic Density}\label{sec:relicdensity}

For the regions of model parameter space considered in this section, dark matter annihilations to standard
model particles proceed via mixing between the dark and ordinary Higgs bosons, often described as the Higgs
portal~\cite{Patt:2006fw}. We take into account mixing between the doublet Higgs fields, $H_D$ and $H$, in our
discussion below. This is consistent with a simplifying assumption that the $\eta$ Higgs does not mix with
the others in the scalar potential.  Such an assumption is adequate for our purposes since we aim only to show
that some parameter region exists in which the correct dark matter relic density is obtained.  Consideration
of a more general potential would likely provide additional solutions in a much larger parameter space, but
would not alter the conclusion that the desired relic density can be achieved.

In this section, $\psi$ will refer to the dark matter mass eigenstate, {\em i.e.}, the lightest mass
eigenstate of the $\psi_u$-$\psi_d$ mass matrix, which we take as diagonal, for convenience.
The potential for the doublet fields has the form:
\begin{equation}
V = -\mu^2 H^\dagger H + \lambda (H^\dagger H)^2 - \mu_D^2 H_D^\dagger H_D
+ \lambda_D (H_D^\dagger H_D)^2 + \lambda_{mix}(H^\dagger H)(H_D^\dagger H_D).
\label{eq:higgsport}
\end{equation}
In unitary gauge, $H$ and $H_D$ are given by
\begin{equation}
H = \frac{1}{\sqrt{2}}\left[\begin{array}{c} 0  \\ v + h \end{array} \right] \, , \hspace{1cm}
H_D = \frac{1}{\sqrt{2}} \left[\begin{array}{c} 0  \\ v_D + h_D \end{array} \right] \, ,
\label{eq:inunitary}
\end{equation}
where $v$ and $v_D$ are the $H$ and $H_D$ vevs, respectively. At the extrema of this potential,
\begin{eqnarray} \label{eq:extremum}
&& v\,(-\mu^2+\lambda \, v^2+\frac{1}{2}\lambda_{mix} \, v_D^2)=0 \nonumber \\
&& v_D \,(-\mu_D^2+\lambda_D \, v_D^2+\frac{1}{2}\lambda_{mix} \, v^2)=0\, .
\end{eqnarray}
The $h$-$h_D$ mass matrix follows from Eq.~(\ref{eq:higgsport}),
\begin{equation} \label{eq:higgsmassmatrix}
M_H^2 = \left( \begin{array}{cc} 2 \, \lambda \, v^2 & \lambda_{mix} \, v\,v_D \\
\lambda_{mix} \, v \, v_D & 2\, \lambda_D \, v_D^2 \end{array} \right).
\end{equation}
Diagonalizing the mass matrix, one finds the mass eigenvalues
\begin{equation}
m_{1,2}^2 = (\lambda_D v_D^2 + \lambda \, v^2) \mp (\lambda_D v_D^2 - \lambda v^2)\sqrt{1+y^2},
\end{equation}
where
\begin{equation}
y=\frac{\lambda_{mix}v \, v_D}{\lambda_D v_D^2 - \lambda \, v^2}.
\end{equation}
The mass eigenstates $h_1$ and $h_2$ are related to $h$ and $h_D$ by a mixing angle
\begin{eqnarray}
h_1 &=& h \cos\theta -  h_D \sin\theta \nonumber \\
h_2 &=& h \sin\theta +  h_D \cos\theta,
\label{eq:hmixing}
\end{eqnarray}
where
\begin{equation}
\tan 2\,\theta = y  \,\,.
\end{equation}

Dark matter annihilations proceed via exchanges of the physical Higgs states $h_1$ and $h_2$.  We take into
account the final states $W^+ W^-$, $Z Z$, $h_1 h_1$ and $t \bar{t}$, where $t$ represents the top quark.
For the parameter choices considered later, final states involving $h_2$ will be subleading. The
relevant annihilation cross sections are given by
\begin{eqnarray} \label{eq:production}
\sigma_{W^+W^-}&=&\frac{g^2 m_{\psi}^2  \sin^2\theta\cos^2\theta}{128 \pi m_W^2 v_D^2} s^2
\left| \frac{1}{s-m_1^2+im_1 \Gamma_1} - \frac{1}{s-m_2^2+im_2\Gamma_2} \right|^2
\nonumber \\ && \times \sqrt{1-\frac{4m_{\psi}^2}{s}} \sqrt{1-\frac{4m_W^2}{s}}
\left(1-\frac{4m_W^2}{s}+\frac{12m_W^4}{s^2}\right) \,\, ,
\label{eq:wwsig}
\end{eqnarray}
\begin{eqnarray}
\sigma_{ZZ}&=&\frac{g^2 m_{\psi}^2 \sin^2\theta\cos^2\theta}{256 \pi m_W^2 v_D^2} s^2
\left| \frac{1}{s-m_1^2+im_1\Gamma_1} - \frac{1}{s-m_2^2+im_2\Gamma_2} \right|^2
\nonumber \\ && \times \sqrt{1-\frac{4m_{\psi}^2}{s}} \sqrt{1-\frac{4m_Z^2}{s}}\left(1-\frac{4m_Z^2}{s}
+\frac{12m_Z^4}{s^2}\right) \,\, ,
\label{eq:zzsig}
\end{eqnarray}
\begin{eqnarray}
\sigma_{h_1h_1}&=&\frac{m_{\psi}^2}{16\pi v_D^2}\left| \frac{3g_{111}\sin\theta}{s-m_1^2+im_1\Gamma_1}
+ \frac{g_{112}\cos\theta}{s-m_2^2+im_2\Gamma_2} \right|^2 \nonumber \\ && \times \sqrt{1-\frac{4m_{\psi}^2}{s}}
\sqrt{1-\frac{4m_{h_1}^2}{s}}\,\, ,
\label{eq:hhsig}
\end{eqnarray}
\begin{eqnarray}
\sigma_{t\bar{t}} &=& \frac{3 m_{\psi}^2 m_t^2 \sin^2\theta\cos^2\theta}{16 \pi v_D^2 v^2} s
\left| \frac{1}{s-m_1^2+im_1 \Gamma_1} - \frac{1}{s-m_2^2+im_2\Gamma_2} \right|^2
\nonumber \\ && \times \left(1-\frac{4 m_t^2}{s}\right) \left(1-\frac{4 m_{\psi}^2}{s}\right) \,\, .
\label{eq:ffsig}
\end{eqnarray}
In Eqs.~(\ref{eq:wwsig}) and (\ref{eq:zzsig}), $g$ is the standard model SU(2) gauge coupling. In
Eq.~(\ref{eq:hhsig}), $g_{111}$ and $g_{112}$ represent the $h_1^3$ and $h_2 h_1^2$ couplings, respectively:
\begin{eqnarray}
g_{111} &=& (\lambda \cos^3\theta + \frac{1}{2} \lambda_{mix} \cos\theta\sin^2\theta)\, v
-(\lambda_D \sin^3\theta+\frac{1}{2} \lambda_{mix} \sin\theta \cos^2\theta)\, v_D  \,\, ,
\nonumber \\
g_{112}  &=&  [3 \lambda \cos^2\theta\sin\theta - \lambda_{mix} (\cos^2\theta\sin\theta-\frac{1}{2} \sin^3\theta)] \, v
\nonumber \\&& \hspace{5em} +\, [3 \lambda_D \sin^2\theta\cos\theta - \lambda_{mix} (\sin^2\theta\cos\theta-\frac{1}{2} \cos^3\theta)]
 \, v_D \,.
\end{eqnarray}
Finally, in all our annihilation cross sections, $\Gamma_1$ ($\Gamma_2$) represents the decay width of the
Higgs field $h_1$ ($h_2$).  The width $\Gamma_1$ is comparable to that of a standard model
Higgs boson and can be neglected without noticeably affecting our numerical results.  However, since
our eventual parameter choices will place the mass of the heavier Higgs field around $2 m_{\psi}$,
we must retain $\Gamma_2$; the leading contributions to $\Gamma_2$ come from the same final
states relevant to the $\psi$ annihilation cross section:
\begin{eqnarray}
\Gamma_{h_2\rightarrow W^+W^-} &=& \frac{g^2m_2^3}{64\pi m_W^2}\sin^2\theta\sqrt{1-\frac{4m_W^2}{m_2^2}}
\left(1-\frac{4m_W^2}{m_2^2}+\frac{12m_W^4}{m_2^4}\right) \nonumber \\
\Gamma_{h_2\rightarrow ZZ} &=& \frac{g^2m_2^3}{128\pi m_W^2}\sin^2\theta\sqrt{1-\frac{4m_Z^2}{m_2^2}}
\left(1-\frac{4m_Z^2}{m_2^2}+\frac{12m_Z^4}{m_2^4}\right) \nonumber \\
\Gamma_{h_2\rightarrow h_1h_1} &=& \frac{g_{112}^2}{32\pi m_2}\sqrt{1-\frac{4m_1^2}{m_2^2}} \nonumber \\
\Gamma_{h_2\rightarrow t \bar{t}} &=& \frac{3 m_2 m_t^2}{8 \pi v^2} \sin^2\theta \left(1-\frac{4 m_t^2}{m_2^2}
\right)^{3/2} \,\, .
\end{eqnarray}\begin{table}
    \centering
        \begin{tabular}{cccccc}
        \hline
        \hline
        $m_\psi$(TeV) & $\sqrt{2\lambda v^2}$(TeV) & $\sqrt{2\lambda_D v_D^2}$(TeV) & $\lambda_{mix}$ & $m_1$(GeV)
        & $m_2$(TeV)
\\
        \hline
        1.0& 0.19& 1.98& 0.21& 158& 1.98\\
        1.5& 0.22& 2.98& 0.28& 199& 2.98\\
        2.0& 0.26& 3.97& 0.39& 241& 3.97\\
        2.5& 0.27& 4.97& 0.42& 257& 4.97\\
        3.0& 0.29& 5.96& 0.52& 277& 5.96\\
        3.5& 0.31& 6.96& 0.57& 299& 6.96\\
        4.0& 0.35& 7.95& 0.70& 339& 7.95\\
        \hline
        \hline
        \end{tabular}
\caption{Examples of viable parameter sets for $v_D = 4$ TeV. For each point
listed, $\Omega_D h^2 \approx 0.1$ and the Higgs masses are consistent with the LEP bound.}
    \label{tab:ParameterSpace}
\end{table}
The evolution of the $\psi$ number density, $n_{\psi}$, is governed by the Boltzmann equation
\begin{equation}
\frac{dn_{\psi}}{dt}+3H(t) n_{\psi} = -\langle \sigma v \rangle [n_{\psi}^2-(n_{\psi}^{EQ})^2],
\end{equation}
where $H(t)$ is the Hubble parameter and $n_{\psi}^{EQ}$ is the equilibrium number density. The
thermally-averaged annihilation cross section times relative velocity $\langle \sigma v \rangle$
is given by~\cite{Gondolo:1990dk}
\begin{equation}
\langle \sigma v \rangle =\frac{1}{8m_\psi^4TK_2^2(m_{\psi}/T)}\int_{4m_{\psi}^2}^\infty
(\sigma_{tot}) \, (s-4m_{\psi}^2)\sqrt{s} \, K_1(\sqrt{s}/T) \, ds \,\,\, ,
\end{equation}
where $\sigma_{tot}$ is the total annihilation cross section, and the $K_i$ are modified Bessel functions
of order $i$.  We evaluate the freeze-out condition~\cite{KolbTurner}
\begin{equation}
\frac{\Gamma}{H(t_F)} \equiv \frac{n_{\psi}^{EQ} \langle \sigma v \rangle}{H(t_F)} \approx 1 \,\, ,
\end{equation}
to find the freeze-out temperature $T_f$, or equivalently $x_f \equiv m_\psi/T_f$. We assume the 
non-relativistic equilibrium number density
\begin{equation}
n_{\psi}^{EQ} = 2 \left(\frac{m_\psi T}{2 \pi}\right)^{3/2} e^{-m_\psi/T} \, ,
\end{equation}
and the Hubble parameter $H=1.66 \,g_*^{1/2} \,T^2/m_{Pl}$, appropriate to a radiation-dominated
universe.  The symbol $g_*$ represents the number of relativistic degrees of freedom and
$m_{Pl}=1.22\times 10^{19}$~GeV is the Planck mass. For the parameter
choices in Tables~\ref{tab:ParameterSpace} and \ref{tab:ParameterSpace2}, we find $x_f \sim 27$--$28$.
We approximate the relic abundance using~\cite{Gondolo:1990dk}
\begin{equation}
\frac{1}{Y_0} = \frac{1}{Y_f} + \sqrt{\frac{\pi}{45}}m_{Pl}m_{\psi}\int_{x_f}^{x_0}
\frac{g_*^{1/2}}{x^2} \langle \sigma v \rangle \, dx
\end{equation}
where $Y$ is the ratio of the number to entropy density and the subscript $0$ indicates the present time.
The ratio of the dark matter relic density to the critical density $\rho_c$ is given by
$\Omega_D = Y_0s_0m_{\psi}/\rho_c$, where $s_0$ is the present entropy density,
or equivalently
\begin{equation}
\Omega_D h^2 \approx 2.8 \times 10^8\mbox{ GeV}^{-1} \, Y_0 \, m_\psi  \,\,\, .
\end{equation}
In our numerical analysis, we assume that the heavy states are sufficiently nondegenerate, so that we do
not have to consider co-annihilation processes~\cite{Griest:1990kh}.  In Tables~\ref{tab:ParameterSpace} and
\ref{tab:ParameterSpace2}, we show representative points in the model's parameter space, spanning a
range of $\psi$ masses, in which we obtain the correct dark matter relic abundance,
$\Omega_D h^2 \approx 0.1$, and in which the masses $m_1$ and $m_2$ are consistent with the 
LEP bound $m_{1,2}>114.4$~GeV~\cite{pdg}.
\begin{table}
    \centering
        \begin{tabular}{cccccc}
        \hline
        \hline
        $m_\psi$(TeV) & $\sqrt{2\lambda v^2}$(TeV) & $\sqrt{2\lambda_D v_D^2}$(TeV) &
$\lambda_{mix}$ & $m_1$(GeV) & $m_2$(TeV)
\\
        \hline
        1.0& 0.16& 1.98& 0.21& 121& 1.98\\
        1.5& 0.15& 2.98& 0.28& 118& 2.98\\
        2.0& 0.16& 3.97& 0.39& 127& 3.97\\
        2.5& 0.15& 4.97& 0.42& 124& 4.97\\
        3.0& 0.15& 5.96& 0.52& 122& 5.96\\
        3.5& 0.15& 6.96& 0.57& 127& 6.96\\
        4.0& 0.15& 7.95& 0.70& 122& 7.95\\
        \hline
        \hline
        \end{tabular}
\caption{Examples of viable parameter sets for $v_D = 4$ TeV, with $m_1$ below $130$~GeV. For each point
listed, $\Omega_D h^2 \approx 0.1$ and the Higgs masses are consistent with the LEP bound.}
\label{tab:ParameterSpace2}
\end{table}
It is common wisdom that weakly interacting dark matter candidates with masses of a few hundred GeV typically
yield relic densities in the correct ballpark.  We have assumed masses above $1$~TeV since most fits to the
positron excess in PAMELA and Fermi LAT indicate that a decaying dark matter candidate should have a mass in this range.
One would therefore expect that $\Omega_D h^2$ in our model should be larger than desirable. The reason this is not
the case is that we have chosen parameters for which the heavier Higgs $h_2$ is within $1\%$ of $2 m_\psi$, leading
to a resonant enhancement in the annihilation rate. While we would be happier without this tuning, it is no larger
than tuning that exists in, for example, the Higgs sector of the minimal supersymmetric standard model.  It is also
worth pointing out that this tuning is related to the portal that connects the dark to standard model sectors of
the theory and is not strictly tied to the mechanism that we have proposed for dark matter decay.  Other portals
are possible.  For example, one might study the limit of the model in which the U(1)$_D$ gauge boson is lighter
and kinetically mixes with hypercharge, a possibility that would lead to other annihilation channels.  Finally, we
point out that Tables~\ref{tab:ParameterSpace} and \ref{tab:ParameterSpace2} includes $m_\psi = 3.5$~TeV, which
naively corresponds to the value preferred by a fit to the PAMELA and Fermi-LAT data, assuming a spin-1/2
dark matter candidate that decays to $\ell^+\ell^-\nu$~\cite{Ibarra:2009dr}.   However, other masses should not
be discounted since astrophysical sources may also contribute to the observed positron excess~\cite{astro}.

\section{Direct Detection}\label{sec:direct}

We now consider whether the parameter choices described in the previous section are consistent
with the current bounds from direct detection experiments.  The most relevant
constraints come from experiments that search for spin-independent, elastic scattering of
dark matter off target nuclei.  The relevant low-energy effective interaction from
$t$-channel exchanges of the Higgs mass eigenstates is given by
\begin{equation}
\mathcal{L}_{int} = \displaystyle\sum_q \alpha_q  \, \bar\psi\psi\, \bar qq \,\, ,
\label{eq:ppqq}
\end{equation}
where
\begin{equation}
\alpha_q = \frac{m_q m_\psi \sin\theta \cos\theta}{v\, v_D} \left(\frac{1}{m^2_1}-\frac{1}{m^2_2}\right).
\end{equation}
\begin{figure}[t]
    \centering
        \includegraphics[width=8cm,angle=0]{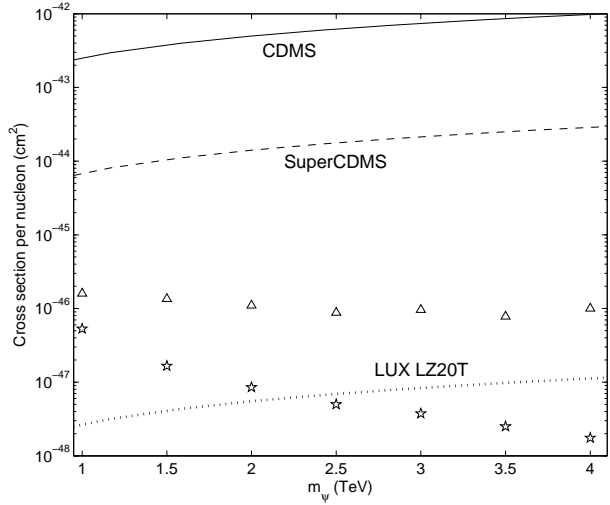}
    \caption{Dark matter-nucleon elastic scattering cross section for the parameter sets in
    Table~\ref{tab:ParameterSpace} (stars) and Table~\ref{tab:ParameterSpace2} (triangles).
The solid line is the current bound from CDMS Soudan 2004-2009 Ge~\cite{Ahmed:2009zw}.
The dashed line represents the projected bound from SuperCDMS Phase A. The dotted line
represents the projected reach of the LUX LZ20T experiment, assuming $1$ event
sensitivity and 13 ton-kilodays.  The graph is obtained using the DM Tools software available at
http://dmtools.brown.edu.}
    \label{fig:direct}
\end{figure}
This interaction is valid for momentum exchanges that are small compared to $m_{1,2}$, which is
always the case given that typical dark matter velocities are non-relativistic.  Following the
approach of Ref.~\cite{Munoz:2003gx}, Eq.~(\ref{eq:ppqq}) leads to an effective interaction with
nucleons
\begin{equation}
\mathcal L_{eff} = f_p  \,\bar\psi\psi \, \bar pp + f_n  \,\bar\psi\psi \, \bar nn \,\, ,
\end{equation}
where $f_p$ and $f_n$ are related to $\alpha_q$ through the relation~\cite{Munoz:2003gx}
\begin{equation}
\frac{f_{p,n}}{m_{p,n}} = \displaystyle\sum_{q=u,d,s} \frac{f^{(p,n)}_{Tq}\alpha_q}{m_q}
+ \frac{2}{27} \, f_{Tg}^{(p,n)}\displaystyle\sum_{q=c,b,t}\frac{\alpha_q}{m_q},
\end{equation}
where $\langle n|m_q\bar qq|n \rangle =m_nf_{Tq}^n$. Numerically, the $f^{(p,n)}_{Tq}$
are given by~\cite{Ellis:2000ds}
\begin{equation}
f_{Tu}^p=0.020\pm0.004,\; f_{Td}^p=0.026\pm0.005,\; f_{Ts}^p = 0.118\pm0.062
\end{equation}
and
\begin{equation}
f_{Tu}^n=0.014\pm0.003,\; f_{Td}^n=0.036\pm0.008,\; f_{Ts}^n = 0.118\pm0.062 \,\, ,
\end{equation}
while $f_{Tg}^{(p,n)}$ is defined by
\begin{equation}
f_{Tg}^{(p,n)}=1-\displaystyle\sum_{q=u,d,s}f_{Tq}^{(p,n)}\,\, .
\end{equation}
We can approximate $f_p \approx f_n$ since $f_{Ts}$ is larger than other $f_{Tq}$'s and $f_{Tg}$. For
the purpose of comparing the predicted cross section with existing bounds, we evaluate the
cross section for scattering off a single nucleon, which can be approximated
\begin{equation}
\sigma_n \approx \frac{m_r^2 f_p^2}{\pi}
\end{equation}
where $m_r$ is nucleon-dark matter reduced mass $1/m_r = 1/m_n+1/m_\psi$.   Our results are shown in
Fig.~\ref{fig:direct}, for the parameter sets given in Tables~\ref{tab:ParameterSpace}
and \ref{tab:ParameterSpace2}. The predicted cross sections are far below the current CDMS
bounds~\cite{Ahmed:2009zw} for dark matter masses between $1$ and $4$~TeV.  However, there is hope that the model
can be probed by the future LUX LZ20T experiment~\cite{lux}.

\section{Conclusions} \label{sec:conc}

We have presented a new TeV-scale model of decaying dark matter.  The approximate
stability of the dark matter candidate, $\psi$, is a consequence of a global U(1) symmetry
that is exact at the perturbative level, but is violated by instanton-induced
interactions of a non-Abelian dark gauge group. The instanton-induced vertex couples
the dark matter candidate to heavy, exotic states that mix with standard model leptons;
the dark matter then decays to $\ell^+ \ell^- \nu$ final states, where the leptons
can be of any generation desired.  We have shown
that a lifetime of $\sim\!10^{26}$~s, which is desirable in decaying dark matter scenarios,
can be obtained for perturbative values of the non-Abelian dark gauge coupling.  In addition,
by studying dark matter annihilations through the Higgs portal, we have provided
examples of parameter regions in which the appropriate dark matter relic density may be
obtained, assuming dark matter masses that are consistent with fits to the results from
the PAMELA and Fermi-LAT experiments. The nucleon-dark matter cross section in our model is
lower than the present bound from CDMS, but may be probed in future experiments.  It might
also be possible to probe the spectrum of our model at the LHC.

The model in this paper provides a concrete, TeV-scale scenario in which dark matter
decay is mediated by instantons, and gives a new motivation for the study of non-Abelian dark gauge
groups~\cite{ndgg}.  However, it is by no means the only possible model of
this type.  One might study variations of the model in which different annihilation channels
are dominant, or the dark matter is lighter, or the standard model leptons are directly
charged under the new non-Abelian gauge group.  It may also be worthwhile to consider how
low-scale leptogenesis and baryogenesis might be accommodated in this type of scenario.  While we have
assumed parameter choices motivated by the observed cosmic ray positron excess,
one might incorporate the present model in a multi-component dark matter scenario if this were 
required to explain new results from ongoing and future direct detection experiments.

\begin{acknowledgments}
This work was supported by the NSF under Grant PHY-0757481. We thank Will Detmold and Marc Sher
for useful comments.
\end{acknowledgments}

\appendix
\section{Mass mixing example}
In Sec.~\ref{sec:model}, we presented a diagrammatic representation of the mixing that
takes the $\chi$ states to standard model leptons.  Here we study the numerical diagonalization
of the corresponding fermion mass matrices, to demonstrate that mixing angles of the size assumed in our
analysis are easily obtained. To simplify the discussion, we focus on mixing with standard model
leptons of a single generation, which we denote by $e$ and $\nu$.  We include (1) Dirac masses
for the $\chi$ fields:
\begin{equation}
{\cal L} \supset \sum_i \left[ a_i \, \overline{\chi}^{(i)}_L \langle H_D \rangle \chi_{uR}^{(i)}
+ b_i \, \overline{\chi}^{(i)}_L \langle H_D \rangle \chi_{dR}^{(i)}
+c_i \, \overline{\chi}^{(i)}_L \langle \widetilde{H}_D \rangle \chi_{uR}^{(i)}
+ d_i \, \overline{\chi}^{(i)}_L \langle \widetilde{H}_D \rangle \chi_{dR}^{(i)} \right] +h.c.\,\,,
\label{eq:cc}
\end{equation}
where $\widetilde{H}_D \equiv i \sigma^2 H^{*}_D$.  These terms generate a completely general two-by-two Dirac
mass matrix for the $\chi$ fermions. (2) Mixing between the $\chi$ fields and standard model leptons:
\begin{eqnarray}
{\cal L} &\supset& g_1 \langle \eta \rangle \overline{\chi}^{(1)}_{dR} e_R^c
+ g_2 \langle \eta \rangle \overline{\chi}^{(1)}_{uR} e_R^c
+ \lambda_e \overline{L} \langle H \rangle e_R \nonumber \\
&+& g_3 \langle \eta \rangle \overline{\chi}^{(2)}_{dR} \nu_R^c
+ g_4 \langle \eta \rangle \overline{\chi}^{(2)}_{uR} \nu_R^c
+ \lambda_\nu \overline{L} \langle \tilde{H} \rangle \nu_R
+ h.c.
\label{eq:csml}
\end{eqnarray}
(3) Mixing involving the vector-like leptons $E_L$ and $E_R$:
\begin{equation}
{\cal L} \supset
g_5 \langle \eta \rangle \, \overline{\chi}^{(3)}_{dR} E_L
+g_6 \langle \eta \rangle \,\overline{\chi}^{(3)}_{uR} E_L
+ M_E \, \overline{E}_L E_R + g_7\, \overline{L} \langle H \rangle E_R +h.c.
\label{eq:vll}
\end{equation}
We now write down the mass matrices which follow from Eqs.~(\ref{eq:cc},\ref{eq:csml},\ref{eq:vll}). For the
neutral states, we work in the basis $f^0_L = (\chi^{(2)}_{uL}, \chi^{(2)}_{dL}, \nu_R^c)$ and
$f^0_R = (\chi^{(2)}_{uR}, \chi^{(2)}_{dR}, \nu_L^c)$.  The neutral mass terms can be written as
$\overline{f^0_L} M_0 f^0_R + h.c.$, where
\begin{equation}
M_0 = \frac{1}{\sqrt{2}} \left(\begin{array}{ccc} c_2 v_D  & d_2 v_D & 0 \\
a_2 v_D & b_2 v_D  & 0 \\ g_4 v_\eta & g_3 v_\eta & \sqrt{2} m_\nu \end{array}
\right) \,\,\, ,
\end{equation}
assuming, for simplicity, that the vevs and couplings are real.  Similarly, the mass terms for the charged
states may be written $\overline{f^-_L} M_c f^-_R + h.c.$, where we assume the basis
$f^-_L=(\chi^{(1)c}_{uR}, \chi^{(1)c}_{dR},\chi^{(3)}_{uL}, \chi^{(3)}_{dL},E_L,
e_L)$ and $f^-_R=(\chi^{(1)c}_{uL}, \chi^{(1)c}_{dL},\chi^{(3)}_{uR}, \chi^{(3)}_{dR},E_R,
e_R)$.  In this case,
\begin{equation}
M_c = \frac{1}{\sqrt{2}} \left(\begin{array}{cccccc}
c_1 v_D & a_1 v_D & 0 & 0 & 0 & g_2 v_\eta \\
d_1 v_D & b_1 v_D & 0 & 0 & 0 & g_1 v_\eta \\
0 & 0 & c_3 v_D & d_3 v_D & 0 & 0 \\
0 & 0 & a_3 v_D & b_3 v_D & 0 & 0 \\
0 & 0 & g_6 v_\eta  &  g_5 v_\eta & \sqrt{2} M_E & 0 \\
0 & 0 & 0 & 0 & g_7 v & \sqrt{2} m_e \end{array}\right) \,\, .
\end{equation}
Given a choice of parameters, it is now a simple matter to compute the relevant
mixing angles numerically.  As an example, let us work in units of the dark scale $v_D$, which
we will assume is $4$~TeV.  In addition we take $v_\eta=v_D$, $M_E=1.5 \,v_D$ and set the standard
model lepton masses to zero (the conclusions do not change if we require realistic standard model
lepton masses).  If one assumes that only the following parameters are nonzero:
\begin{eqnarray}
&&\{b_1,c_1,b_2,c_2,b_3,c_3,g_1,g_2,g_3,g_4,g_5,g_6,g_7\}= \nonumber\\
&&\{1.9,1.8,1.8,1.7,2.1,2.0,0.02,0.02,0.02,0.02,0.7,0.6,1.0\}  \,\,,
\end{eqnarray}
then one finds
\begin{center}
\begin{tabular}{lcr}
$\chi^{(1)}_{uL}=0.011 \, {e_R^c}_{0} + \cdots$ & \qquad\qquad &$\chi^{(1)}_{dL} = 0.011 \, {e_R^c}_{0} + \cdots$ \\
$\chi^{(2)}_{uL}=  0.012 \, {\nu_R^c}_{0} + \cdots$ & &$\chi^{(2)}_{dL} = 0.011 \, {\nu_R^c}_{0} + \cdots$\\
$\chi^{(3)}_{uL} = 0.009 \, {e_L}_{0} + \cdots$ & &$\chi^{(3)}_{dL} = 0.010 \, {e_L}_{0} + \cdots$ \\
\end{tabular} 
\end{center}
where the fields on the right represent mass eigenstates.  In addition, the non-zero mass eigenvalues are
all larger than the $\psi$ mass if $m_\psi < 1.2 \,v_D$, so that only decays to standard model
leptons via the instanton vertex are kinematically allowed. Given the number of free
parameters involved, one sees that the mixing angles are highly model dependent and can be easily set
to the values assumed in Sec.~\ref{sec:model}.


\begin{thebibliography}{99}

\bibitem{Abdo:2009zk}
  A.~A.~Abdo {\it et al.}  [The Fermi LAT Collaboration],
  Phys.\ Rev.\ Lett.\  {\bf 102}, 181101 (2009)
  [arXiv:0905.0025 [astro-ph.HE]].

\bibitem{Aharonian:2009ah}
  F.~Aharonian {\it et al.}  [H.E.S.S. Collaboration],
  Astron.\ Astrophys.\  {\bf 508}, 561 (2009)
  [arXiv:0905.0105 [astro-ph.HE]].

\bibitem{Adriani:2008zr}
  O.~Adriani {\it et al.}  [PAMELA Collaboration],
  Nature {\bf 458}, 607 (2009)
  [arXiv:0810.4995 [astro-ph]].

\bibitem{Adriani:2008zq}
  O.~Adriani {\it et al.},
  Phys.\ Rev.\ Lett.\  {\bf 102}, 051101 (2009)
  [arXiv:0810.4994 [astro-ph]].

\bibitem{astro}
  D.~Hooper, P.~Blasi and P.~D.~Serpico,
  JCAP {\bf 0901}, 025 (2009)
  [arXiv:0810.1527 [astro-ph]];
 H.~Yuksel, M.~D.~Kistler and T.~Stanev,
  Phys.\ Rev.\ Lett.\  {\bf 103}, 051101 (2009)
  [arXiv:0810.2784 [astro-ph]].

\bibitem{Cholis:2008hb}
  I.~Cholis, L.~Goodenough, D.~Hooper, M.~Simet and N.~Weiner,
  Phys.\ Rev.\  D {\bf 80}, 123511 (2009)
  [arXiv:0809.1683 [hep-ph]];
  M.~Cirelli, M.~Kadastik, M.~Raidal and A.~Strumia,
  Nucl.\ Phys.\  B {\bf 813}, 1 (2009)
  [arXiv:0809.2409 [hep-ph]].

\bibitem{ArkaniHamed:2008qn}
  N.~Arkani-Hamed, D.~P.~Finkbeiner, T.~R.~Slatyer and N.~Weiner,
  Phys.\ Rev.\  D {\bf 79}, 015014 (2009)
  [arXiv:0810.0713 [hep-ph]].


\bibitem{MarchRussell:2008tu}
  J.~D.~March-Russell and S.~M.~West,
  Phys.\ Lett.\  B {\bf 676}, 133 (2009)
  [arXiv:0812.0559 [astro-ph]].

\bibitem{Pospelov:2008jd}
  M.~Pospelov and A.~Ritz,
  Phys.\ Lett.\  B {\bf 671}, 391 (2009)
  [arXiv:0810.1502 [hep-ph]].

\bibitem{feldman}
  D.~Feldman, Z.~Liu and P.~Nath,
  Phys.\ Rev.\  D {\bf 79}, 063509 (2009)
  [arXiv:0810.5762 [hep-ph]].

\bibitem{Ibe:2008ye}
  M.~Ibe, H.~Murayama and T.~T.~Yanagida,
  Phys.\ Rev.\  D {\bf 79}, 095009 (2009)
  [arXiv:0812.0072 [hep-ph]].

\bibitem{Guo:2009aj}
  W.~L.~Guo and Y.~L.~Wu,
  Phys.\ Rev.\  D {\bf 79}, 055012 (2009)
  [arXiv:0901.1450 [hep-ph]].

\bibitem{ddecay}
W.~L.~Guo, Y.~L.~Wu and Y.~F.~Zhou,
Phys.\ Rev.\  D {\bf 81}, 075014 (2010)
[arXiv:1001.0307 [hep-ph]];
  M.~Luo, L.~Wang, W.~Wu and G.~Zhu,
  Phys.\ Lett.\  B {\bf 688}, 216 (2010)
  [arXiv:0911.3235 [hep-ph]];
  B.~Kyae,
  Phys.\ Lett.\  B {\bf 685}, 19 (2010)
  [arXiv:0909.3139 [hep-ph]];
  J.~H.~Huh and J.~E.~Kim,
  Phys.\ Rev.\  D {\bf 80}, 075012 (2009)
  [arXiv:0908.0152 [hep-ph]];
  M.~R.~Buckley, K.~Freese, D.~Hooper, D.~Spolyar and H.~Murayama,
  Phys.\ Rev.\  D {\bf 81}, 016006 (2010)
  [arXiv:0907.2385 [astro-ph.HE]];
  D.~Aristizabal Sierra, D.~Restrepo and O.~Zapata,
  Phys.\ Rev.\  D {\bf 80}, 055010 (2009)
  [arXiv:0907.0682 [hep-ph]];
  C.~H.~Chen, C.~Q.~Geng and D.~V.~Zhuridov,
  JCAP {\bf 0910}, 001 (2009)
  [arXiv:0906.1646 [hep-ph]];
  A.~Arvanitaki, S.~Dimopoulos, S.~Dubovsky, P.~W.~Graham, R.~Harnik and S.~Rajendran,
  Phys.\ Rev.\  D {\bf 80}, 055011 (2009)
  [arXiv:0904.2789 [hep-ph]];
  S.~L.~Chen, R.~N.~Mohapatra, S.~Nussinov and Y.~Zhang,
  Phys.\ Lett.\  B {\bf 677}, 311 (2009)
  [arXiv:0903.2562 [hep-ph]];
  K.~Ishiwata, S.~Matsumoto and T.~Moroi,
  JHEP {\bf 0905}, 110 (2009)
  [arXiv:0903.0242 [hep-ph]];
  S.~Shirai, F.~Takahashi and T.~T.~Yanagida,
  arXiv:0902.4770 [hep-ph];
  X.~Chen,
  JCAP {\bf 0909}, 029 (2009)
  [arXiv:0902.0008 [hep-ph]];
  C.~H.~Chen, C.~Q.~Geng and D.~V.~Zhuridov,
  Phys.\ Lett.\  B {\bf 675}, 77 (2009)
  [arXiv:0901.2681 [hep-ph]];
  K.~Hamaguchi, F.~Takahashi and T.~T.~Yanagida,
  Phys.\ Lett.\  B {\bf 677}, 59 (2009)
  [arXiv:0901.2168 [hep-ph]];
  E.~Nardi, F.~Sannino and A.~Strumia,
  JCAP {\bf 0901}, 043 (2009)
  [arXiv:0811.4153 [hep-ph]];
  C.~R.~Chen, M.~M.~Nojiri, F.~Takahashi and T.~T.~Yanagida,
  Prog.\ Theor.\ Phys.\  {\bf 122}, 553 (2009)
  [arXiv:0811.3357 [astro-ph]];
  K.~Hamaguchi, E.~Nakamura, S.~Shirai and T.~T.~Yanagida,
  Phys.\ Lett.\  B {\bf 674}, 299 (2009)
  [arXiv:0811.0737 [hep-ph]];
  P.~f.~Yin, Q.~Yuan, J.~Liu, J.~Zhang, X.~j.~Bi and S.~h.~Zhu,
  Phys.\ Rev.\  D {\bf 79}, 023512 (2009)
  [arXiv:0811.0176 [hep-ph]];
  C.~R.~Chen, F.~Takahashi and T.~T.~Yanagida,
  Phys.\ Lett.\  B {\bf 671}, 71 (2009)
  [arXiv:0809.0792 [hep-ph]];
  I.~Gogoladze, R.~Khalid, Q.~Shafi and H.~Yuksel,
  Phys.\ Rev.\  D {\bf 79}, 055019 (2009)
  [arXiv:0901.0923 [hep-ph]].




\bibitem{Ibarra:2008jk}
  A.~Ibarra and D.~Tran,
  JCAP {\bf 0902}, 021 (2009)
  [arXiv:0811.1555 [hep-ph]].

\bibitem{Ibarra:2009dr}
  A.~Ibarra, D.~Tran and C.~Weniger,
  JCAP {\bf 1001}, 009 (2010)
  [arXiv:0906.1571 [hep-ph]].

\bibitem{Kuzmin:1997cm}
  V.~A.~Kuzmin and V.~A.~Rubakov,
  Phys.\ Atom.\ Nucl.\  {\bf 61}, 1028 (1998)
  [Yad.\ Fiz.\  {\bf 61}, 1122 (1998)]
  [arXiv:astro-ph/9709187].

\bibitem{kubo}
  H.~Fukuoka, J.~Kubo and D.~Suematsu,
  Phys.\ Lett.\  B {\bf 678}, 401 (2009)
  [arXiv:0905.2847 [hep-ph]].

\bibitem{thooft1}
 G.~'t Hooft,
  Phys.\ Rev.\ Lett.\  {\bf 37}, 8 (1976).

\bibitem{thooft2}
  G.~'t Hooft,
  Phys.\ Rev.\  D {\bf 14}, 3432 (1976)
  [Erratum-ibid.\  D {\bf 18}, 2199 (1978)].

\bibitem{noble}
 R.~J.~Noble,
  Phys.\ Rev.\  D {\bf 25}, 825 (1982).

\bibitem{Patt:2006fw}
  B.~Patt and F.~Wilczek,
  arXiv:hep-ph/0605188.

\bibitem{Gondolo:1990dk}
  P.~Gondolo and G.~Gelmini,
  Nucl.\ Phys.\  B {\bf 360}, 145 (1991).

\bibitem{KolbTurner}
    E. W. Kolb and M. S. Turner, "The Early Universe," \textit{Boulder, Colorado: Westview Press (1994) 547p.}

\bibitem{Griest:1990kh}
  K.~Griest and D.~Seckel,
  Phys.\ Rev.\  D {\bf 43}, 3191 (1991).

\bibitem{pdg}
 C.~Amsler {\it et al.}  [Particle Data Group],
  Phys.\ Lett.\  B {\bf 667}, 1 (2008).

\bibitem{Ahmed:2009zw}
  Z.~Ahmed {\it et al.}  [The CDMS-II Collaboration],
  arXiv:0912.3592 [astro-ph.CO].

\bibitem{Munoz:2003gx}
  C.~Munoz,
  Int.\ J.\ Mod.\ Phys.\  A {\bf 19}, 3093 (2004)
  [arXiv:hep-ph/0309346].

\bibitem{Ellis:2000ds}
  J.~R.~Ellis, A.~Ferstl and K.~A.~Olive,
  Phys.\ Lett.\  B {\bf 481}, 304 (2000)
  [arXiv:hep-ph/0001005].

\bibitem{lux}
  D.~N.~McKinsey {\it et al.},
  J.\ Phys.\ Conf.\ Ser.\  {\bf 203}, 012026 (2010);
  S.~Fiorucci {\it et al.},
  AIP Conf.\ Proc.\  {\bf 1200}, 977 (2010)
  [arXiv:0912.0482 [astro-ph.CO]].

\bibitem{ndgg}
  J.~L.~Diaz-Cruz and E.~Ma,
  arXiv:1007.2631 [hep-ph];
  H.~Zhang, C.~S.~Li, Q.~H.~Cao and Z.~Li,
  arXiv:0910.2831 [hep-ph];
  F.~Chen, J.~M.~Cline and A.~R.~Frey,
  Phys.\ Rev.\  D {\bf 80}, 083516 (2009)
  [arXiv:0907.4746 [hep-ph]];
  M.~Baumgart, C.~Cheung, J.~T.~Ruderman, L.~T.~Wang and I.~Yavin,
  JHEP {\bf 0904}, 014 (2009)
  [arXiv:0901.0283 [hep-ph]];
  D.~G.~E.~Walker,
  arXiv:0907.3146 [hep-ph].
\end{thebibliography}
\end{document}